\documentclass[journal]{IEEEtran}

\usepackage{cite}

\ifCLASSINFOpdf
\usepackage[pdftex]{graphicx}
\graphicspath{{../pdf/}{../jpeg/}}
\DeclareGraphicsExtensions{.pdf,.jpeg,.png}
\else
\fi
\UseRawInputEncoding
\usepackage{amsmath}
\usepackage{array}
\usepackage{standalone}
\usepackage{etoolbox}
\newtoggle{arXiv}%\toggletrue{arXiv}
\newtoggle{bibrun}%\toggletrue{bibrun}
\usepackage{newtxtext, newtxmath}
\usepackage{graphicx}
\usepackage{color, calc}
\usepackage{array, booktabs}
\usepackage{tikz, tikz-3dplot}
\usetikzlibrary{arrows, arrows.meta, calc, positioning, decorations.pathmorphing}
\tikzset{>=Stealth}
\usepackage{siunitx}

\usepackage{enumitem}
\setlist[description]{labelindent=0pt, leftmargin=\parindent, font=\normalfont\itshape}
\usepackage{url}
\usepackage{subfigure}
\usepackage{textcomp}
\usepackage{eurosym}
\usepackage{xfrac}
\usepackage{pgfplots}
\usepackage{stmaryrd}
\pgfplotsset{compat=1.17}
\usepackage{makecell}
\usepackage[utf8]{inputenc}
\usepackage{hyperref}
\begin{document}
	
	\title{Comment on electromagnetic noise cancellation in low-field MRI systems \ (arXiv:2509.05955v1, 2406.17804v3, 2210.06730v2, and related works)}
	
	\author{\IEEEauthorblockN{
			Joseba~Alonso\IEEEauthorrefmark{1},
			Jos\'e~M.~Algar\'in\IEEEauthorrefmark{1}$^,$\IEEEauthorrefmark{2},
			and~Teresa~Guallart-Naval\IEEEauthorrefmark{1}}
		
		\IEEEauthorblockA{\IEEEauthorrefmark{1}MRILab, Institute for Molecular Imaging and Instrumentation (i3M), Consejo Superior de Investigaciones Cient\'ificas (CSIC) \& Universitat Polit\`ecnica de Val\`encia (UPV), Valencia, Spain}\\
		\IEEEauthorblockA{\IEEEauthorrefmark{2}Full Body Insight. S.L., Paterna, Spain}
		
		\thanks{Corresponding author: J. Alonso (joseba.alonso@i3m.upv.es).}
	}
	
	\maketitle
	
	\begin{abstract}
		\newline
		In this Comment, we discuss recent approaches to electromagnetic interference (EMI) mitigation in low-field Magnetic Resonance Imaging (LF-MRI), as presented in arXiv preprints \href{https://arxiv.org/abs/2509.05955}{2509.05955},
		\href{https://arxiv.org/abs/2406.17804}{2406.17804}, or
		\href{https://arxiv.org/abs/2210.06730}{2210.06730}. These and other works explore noise cancellation strategies based on external sensing coils for post-elimination of EMI. We argue that, under realistic conditions, such approaches lead to residual signal contamination that necessarily exceed that obtained with optimal hardware-based pre-elimination.
	\end{abstract}
	
	\IEEEpeerreviewmaketitle

This Comment addresses recent work on electromagnetic interference (EMI) mitigation in Low-Field Magnetic Resonance Imaging (LF-MRI), embodied to a large extent in arXiv preprints \href{https://arxiv.org/abs/2509.05955}{2509.05955v1} \cite{He2025},
\href{https://arxiv.org/abs/2406.17804}{2406.17804v3} \cite{Bian2024} (peer-reviewed version in Ref.~\cite{Bian2024b}), and
\href{https://arxiv.org/abs/2210.06730}{2210.06730v2} \cite{Zhao2023} (peer-reviewed version in Ref.~\cite{Zhao2024b}). These studies propose noise cancellation (NC) strategies based on auxiliary sensing coils to estimate and remove EMI contributions from the receive signal (see Refs.~\cite{Srinivas2022,Zhao2024b,Wang2026} and references therein).

In NC approaches, EMI is inferred from signals measured on external sensing coils and subsequently removed through post-processing. This strategy has shown promising reductions of structured interference in reconstructed images and is particularly attractive for portable LF-MRI systems.

Our point of departure is that these approaches implicitly assume that the noise contributions captured by sensing coils suffice to characterize the noise at the receive coil. In the following, we examine this assumption and its implications for noise and signal-to-noise ratio (SNR) quantification.

LF-MRI scanners are often operated in unshielded environments, making them susceptible to EMI pick-up on the radio-frequency (RF) coils used for signal detection~\cite{webb2023b}. Importantly, MRI signal detection is fundamentally a near-field magnetic phenomenon~\cite{Hoult1989}, with receive coils designed to couple to time-varying magnetic flux rather than to propagating electromagnetic waves \cite{Brunner2009}. As a result, to a really good approximation, external interference that couples efficiently into the receive chain does not arise from the same inductive mechanisms responsible for signal detection, but instead involves alternative pathways, predominantly electric-field interactions with the subject and coil~\cite{Pfitzer2025,He2026}.

EMI can be mitigated either through hardware-based pre-elimination (e.g., RF grounding and shielding~\cite{GuallartNaval2026a}) or through post-processing strategies such as NC (see Ref.~\cite{Wang2026} and references therein). However, assessing the effectiveness of these approaches requires unambiguous noise quantification.

Fortunately, this need not be addressed in the unitless space of image reconstructions. Rather, MRI signals are electromagnetic in nature, their noise can be quantified in volts, and this can be benchmarked against the absolute minimum noise achievable which, at low fields, corresponds to thermal Johnson noise\cite{GuallartNaval2026a}.

Here we argue that NC approaches invariably result in stronger signal contamination than pre-elimination based on optimal hardware noise suppression. This may have gone unnoticed because, to our best knowledge, no post-elimination papers quantify noise in volts or benchmark it against the thermal limit.

NC techniques rely on the determination of the filter response or transfer functions that characterize the relation between the EMI picked up in one or various \emph{sensing} coils and the EMI at the \emph{receive} antennae. For the sake of simplicity, we will hereon assume a single receive coil, but the arguments in this Comment trivially generalize to multiple receptors.

The key concepts underlying NC are: i) the time-varying voltage on the receive coil during a readout is
\begin{equation}\label{eq:v0}
    v_0(t) = s_\text{MR}(t) + n_0(t),
\end{equation}
where $s_\text{MR}(t)$ is the magnetic resonance signal that contains relevant information and $n_0(t)$ is the confounding noise contribution; ii) the voltages on the sensing coils result exclusively from noise contributions
\begin{equation}\label{eq:vi}
    v_i(t) = n_i(t) \text{, for } 1\leq i\leq N,
\end{equation}
with $N$ the number of sensing coils available; and iii) the noise at the sensing and receive coils are related by transfer functions $h_i(t)$ as
\begin{equation}\label{eq:TF}
    n_0(t) = \sum_i h_i(t) * n_i(t).
\end{equation}

Prior to any data processing, the signal-to-noise ratio (SNR) at the receive coil is
\begin{equation}\label{eq:snr}
    \text{SNR} = \sqrt{\frac{\sigma_{s_\text{MR}}^2}{\sigma_{n_0}^2}},
\end{equation}
where $\sigma_{x}^2$ is the variance of $x$ and we have removed the explicit time dependencies. However, Eq.~(\ref{eq:TF}) implies that $n_0$ can be unambiguously determined from $n_i$, given that $h_i$ are known. Ultimately, $h_i$ characterize the differences in electromagnetic paths followed by the EMI towards the different coils. If they are indeed known, $n_0$ can be post-eliminated by subtraction from $v_0$,
\begin{equation}\label{eq:v0p}
    v'_0 = v_0 - \sum_i h_i*n_i = s_\text{MR},
\end{equation}
and the SNR of $v'_0$ would be infinite. 

Determining the $h_i$ functions from first principles is challenging. Instead, there have appeared a number of methods that estimate $h_i$ following multiple acquisitions with $N\geq1$, be them analytical (e.g. \cite{Wang2026,Srinivas2022}) or machine-learning-based algorithms (e.g. \cite{Zhao2024b}).

The question that begs is then: \emph{in practice, why can we not wholly post-eliminate receive noise based on readouts on a sensing coil?} The answer is in \textbf{uncorrelated noise}.

To account for uncorrelated noise, Eqs.~(\ref{eq:v0}), (\ref{eq:vi}), and (\ref{eq:snr}) can be rewritten as
\begin{equation}\label{eq:v02}
    v_0 = s_\text{MR} + n_{0,\text{coil}} + n_{0,\text{load}} + n_{0,\text{emi}},
\end{equation}
\begin{equation}\label{eq:vi2}
    v_i = n_{i,\text{coil}} + n_{i,\text{emi}} \text{, for } 1\leq i\leq N,
\end{equation}
and
\begin{equation}\label{eq:snr2}
    \text{SNR} = \sqrt{\frac{\sigma_{s_\text{MR}}^2}{\sigma_{n_{0,\text{coil}}}^2 + \sigma_{n_{0,\text{load}}}^2 + \sigma_{n_{0,\text{emi}}}^2}},
\end{equation}
where $n_\text{coil}$ refers to intrinsic (thermal) noise in the coil, $n_\text{load}$ refers to noise generated within the sample/subject and coupled to the receive coil \cite{Marques2019}, and $n_\text{emi}$ refers to external EMI picked up by both the receive and sensing coils. The latter can be correlated between the coils through transfer functions $h_i$; $n_\text{coil}$ and $n_\text{load}$ feature no such correlations.

All things considered, the NC procedures described above not only do not post-eliminate all noise contributions; they actually transfer the intrinsic noise of the sensing coils into the receive coil according to $h_i$. In the best possible scenario, Eq.~(\ref{eq:TF}) becomes
\begin{equation}\label{eq:TF2}
    n_{0,\text{emi}} = \sum_i h_i * n_{i,\text{emi}},
\end{equation}
Eq.~(\ref{eq:v0p}) results in
\begin{eqnarray}\label{eq:v0p2}
    \nonumber v'_0 & = & v_0 - \sum_i h_i*n_i \\ 
    & = & s_\text{MR} + n_{0,\text{coil}} + n_{0,\text{load}} - \sum_i h_i * n_{i,\text{coil}},
\end{eqnarray}
and
\begin{equation}\label{eq:snr3}
    \text{SNR} = \sqrt{\frac{\sigma_{s_\text{MR}}^2}{\sigma_{n_{0,\text{coil}}}^2 + \sigma_{n_{0,\text{load}}}^2 + \sum_i \sigma_{h_i * n_{i,\text{coil}}}^2}}.
\end{equation}
This can be worse than Eq.~(\ref{eq:snr2}) and is invariably lower than the case where EMI is pre-eliminated with proper RF grounding with conductive blankets \cite{Guallart-Naval2022}, direct subject grounding \cite{Lena2026}, or by removing the capacitive coupling between the subject and receive coils \cite{Pfitzer2025} (while retaining the inductive coupling which is to a good approximation irrelevant in terms of EMI pickup). For these cases,
\begin{equation}\label{eq:snr4}
    \text{SNR} = \sqrt{\frac{\sigma_{s_\text{MR}}^2}{\sigma_{n_{0,\text{coil}}}^2 + \sigma_{n_{0,\text{load}}}^2}}.
\end{equation}
Furthermore, LF experiments with good RF grounding demonstrate that the total noise can be made compatible with Johnson thermal levels, i.e.
\begin{equation}
    \sigma_{n_{0,\text{load}}}^2 \ll \sigma_{n_{0,\text{coil}}}^2 \Rightarrow \text{SNR} \approx \sqrt{\frac{\sigma_{s_\text{MR}}^2}{\sigma_{n_{0,\text{coil}}}^2}},
\end{equation}
even in massively EMI-contaminated environments \cite{Algarin2023}.

Post-elimination may still be advantageous in certain scenarios, such as in absence of quality grounding material or in rare cases of inductively-coupled EMI. However, this does not remove the need for unambiguous noise reporting (i.e. in volts, not as the fraction of EMI removed by one or another technique) in future LF-MRI publications.

%\bibliography{myrefs}

\end{document}